\documentclass[12pt,,letterpaper]{JHEP3}
\usepackage{graphics}
\usepackage{epsfig}

\title{$SL(2,R)$ symmetry and quasi-normal modes in the BTZ black hole}

\author{Hongbao Zhang\\
Crete Center for Theoretical Physics, Department of Physics, \\
University of Crete, 71003 Heraklion, Greece\\
\email{hzhang@physics.uoc.gr}}

\abstract{With the help of two new intrinsic tensor fields
associated with the $SL(2,R)$ quadratic Casimir of Killing fields,
 we uncover the $SL(2,R)$ symmetry satisfied by the
solutions to the equations of motion for various fields in the BTZ
black hole in a uniform way by performing tensor and spinor analysis
without resorting to any specific coordinate system. Then with the
standard algebraic method developed recently, we determine the
quasi-normal modes for various fields in the BTZ black hole. As a
result, the quasi-normal modes are given by the infinite tower of
descendants of the chiral highest weight mode, which is in good
agreement with the previous analytic result obtained by exactly
solving equations of motion instead.}

\begin{document}
\section{Introduction}
As a concrete implementation of holographic principle, AdS/CFT
correspondence is definitely one of the most significant
developments in fundamental physics. With this holographic
dictionary, the black hole in the bulk is dual to the conformal
field theory at a finite temperature. In particular, as first
suggested and numerically tested in \cite{HH}, the quasi-normal
modes associated with the linear perturbation of the black hole
correspond to the operators perturbing the dual field theory at the
thermal equilibrium, and the quasi-normal frequencies arise as the
poles of the retarded Green function of the operators. Due to the
simplicity of AdS$_3$/CFT$_2$, where exact computations can be
performed on both sides, the precise quantitative agreement has been
confirmed for scalar, fermion, and vector perturbations of the BTZ
black hole by solving the equations of motion
analytically\cite{BSS}. Later, in the spirit of \cite{MS} and
\cite{BKL}, it is demonstrated that the quasi-normal modes can be
constructed as an infinite tower of descendants of left and right
chiral highest weight mode associated with the $SL(2,R)$ Lie algebra
of Killing fields for scalar and tensor perturbations of the BTZ
black hole\cite{SS}. More recently, such an algebraic construction
has been generalized to other three dimensional black holes with the
vector perturbation included\cite{CL,CZ}. However, the intermediate
calculation involved there is a little bit complicated, which makes
the $SL(2,R)$ symmetry lost in our eyes although it is uncovered in
the final step somehow. The technical reason for that comes
partially from the fact that the ordinary derivative is used there
rather than the covariant derivative. It is highly expected that the
computation will be simplified once one adopts the covariant
derivative instead since it is intrinsic to the spacetime geometry.

The purpose of this paper is two fold. One is to develop the
simplified computational strategy by making transparent the
$SL(2,R)$ symmetry during the course of our analysis. To achieve
this, besides appealing to the covariant derivative as alluded to
above, we also introduce two intrinsic tensor fields associated with
the $SL(2,R)$ symmetry. It turns out that these two guys have very
nice properties and always conspire to organize and manipulate the
calculation as simple as possible. The other is to incorporate the
fermion field into our analysis in a uniform way.

The rest of paper is structured as follows. In the next section,
after a brief review of the $SL(2,R)$ Lie algebra of Killing fields
in the BTZ black hole and its quadratic Casimir operator, we
introduce the two intrinsic tensor fields associated with the
quadratic Casimir and uncover their intriguing features. With this
preparation, we provide an explicit derivation of how the solutions
to equations of motion fall into the representation of the $SL(2,R)$
Lie algebra for various fields in the BTZ black hole in Section
\ref{derivation}. Then in the subsequent section, we construct the
corresponding quasi-normal modes by the standard algebraic approach.
As expected, the result is in good agreement with the previous
computation. The conclusion and discussion are put into the last
section.

Notation and conventions follow \cite{Wald} unless specified
otherwise.
\section{$SL(2,R)$ quadratic Casimir in the BTZ black hole}
\subsection{Two sets of $SL(2,R)$ Lie algebra associated with the Killing
fields}
Without loss of generality, let us start with the non-rotating BTZ
black hole with unit mass, i.e.,
\begin{equation}
ds^2=-\sinh^2(\rho)d\tau^2+\cosh^2(\rho)d\varphi^2+d\rho^2.
\end{equation}
In what follows, we would like to work in the light cone
coordinates, i.e., $u=\tau+\varphi, v=\tau-\varphi$, in which the
metric takes the form
\begin{equation}
g_{ab}=\frac{1}{4}\{(du)_a(du)_b-\cosh(2\rho)[(du)_a(dv)_b+(du)_b(dv)_a]+(dv)_a(dv)_b\}+(d\rho)_a(d\rho)_b.
\end{equation}
Whence the inverse metric can be obtained as
\begin{eqnarray}
g^{ab}&=&-\frac{4}{\sinh^2(2\rho)}\{(\frac{\partial}{\partial
u})^a(\frac{\partial}{\partial
u})^b+\cosh(2\rho)[(\frac{\partial}{\partial
u})^a(\frac{\partial}{\partial v})^b+(\frac{\partial}{\partial
u})^b(\frac{\partial}{\partial
v})^a]\nonumber\\
&&+(\frac{\partial}{\partial v})^a(\frac{\partial}{\partial
v})^a\}+(\frac{\partial}{\partial\rho})^a(\frac{\partial}{\partial\rho})^b,
\end{eqnarray}
and the associated volume element reads
\begin{equation}
\epsilon=\frac{\sinh(2\rho)}{4}du\wedge dv\wedge d\rho.
\end{equation}

Note that the above BTZ black hole is locally $AdS_3$. In
particular, the Riemann and Ricci tensors are given by
\begin{equation}
R_{abcd}=g_{ad}g_{bc}-g_{ac}g_{bd},R_{ab}=-2g_{ab}.
\end{equation}
Thus such a black hole also admits six Killing fields. A Killing
field $\xi$, by definition, is a vector field which can generate
one-parameter group of isometries, or equivalently, a vector field
satisfying the Killing equation $\nabla_a\xi_b=\nabla_{[a}\xi_{b]}$
with $\nabla_a$ the covariant derivative operator. With this, one
can easily show that the Lie derivative with respect to any Killing
field Kills all intrinsic tensor fields associated with the metric
such as the volume element and commutes with the covariant
derivative operator. Here we denote these six Killing fields by
$L_k$ and $\bar{L}_k$ with $k=0,\pm1$. In particular, $L_k$ is given
by
\begin{eqnarray}
L_0^a&=&-(\frac{\partial}{\partial u})^a, \nonumber\\
L_{-1}^a&=&e^{-u}[-\frac{\cosh(2\rho)}{\sinh(2\rho)}(\frac{\partial}{\partial
u})^a-\frac{1}{\sinh(2\rho)}(\frac{\partial}{\partial
v})^a-\frac{1}{2}(\frac{\partial}{\partial\rho})^a],\nonumber\\
L_{+1}^a&=&e^{u}[-\frac{\cosh(2\rho)}{\sinh(2\rho)}(\frac{\partial}{\partial
u})^a-\frac{1}{\sinh(2\rho)}(\frac{\partial}{\partial
v})^a+\frac{1}{2}(\frac{\partial}{\partial\rho})^a].\label{Killing}
\end{eqnarray}
Similarly, $\bar{L}_k$ is defined as (\ref{Killing}) simply by
switching $u$ and $v$ therein. Locally their Lie commutators satisfy
two sets of the $SL(2,R)$ Lie algebra, i.e.,
\begin{equation}
[L_0,L_{\pm1}]=\mp L_{\pm1},[L_{+1},L_{-1}]=2L_0,
[\bar{L}_0,\bar{L}_{\pm1}]=\mp
\bar{L}_{\pm1},[\bar{L}_{+1},\bar{L}_{-1}]=2\bar{L}_0.
\end{equation}
Note that the Lie derivative conforms to
$[\mathcal{L}_X,\mathcal{L}_Y]=\mathcal{L}_{[X,Y]}$ and
$\mathcal{L}_{\alpha X}=\alpha\mathcal{L}_{X}$ for the arbitrary
vector fields $X$ and $Y$ with the arbitrary constant
$\alpha$\footnote{When the Lie derivative acts on the spinor, the
situation will become a little bit subtle. In particular,
$[\mathcal{L}_X,\mathcal{L}_Y]=\mathcal{L}_{[X,Y]}$ holds only for
the case in which either $X$ or $Y$ is the conformal Killing
field\cite{CD}. Fortunately this subtlety does not bother us as we
focus only on the Lie derivative along Killing fields here.} Thus
the above Lie algebra can be naturally represented by the Lie
derivative. In particular, the quadratic Casimir operators can be
realized by the Lie derivative as
\begin{equation}
\mathcal{L}^2=\mathcal{L}_{L_0}\mathcal{L}_{L_0}-\frac{1}{2}(\mathcal{L}_{L_{+1}}\mathcal{L}_{L_{-1}}+\mathcal{L}_{L_{-1}}\mathcal{L}_{L_{+1}}),
\bar{\mathcal{L}}^2=\mathcal{L}_{\bar{L}_0}\mathcal{L}_{\bar{L}_0}-\frac{1}{2}(\mathcal{L}_{\bar{L}_{+1}}\mathcal{L}_{\bar{L}_{-1}}+\mathcal{L}_{\bar{L}_{-1}}\mathcal{L}_{\bar{L}_{+1}}),\label{commutation}
\end{equation}
which commute with $\mathcal{L}_{L_k}$ and
$\mathcal{L}_{\bar{L}_k}$.
\subsection{Two types of tensor fields associated with the quadratic Casimir}
Now we would like to construct the two types of tensor fields
associated with the quadratic $SL(2,R)$ Casimir as follows
\begin{equation}
H^{ab}=L_0^aL_0^b-\frac{1}{2}(L_{+1}^aL_{-1}^b+L_{-1}^aL_{+1}^b),\bar{H}^{ab}=\bar{L}_0^a\bar{L}_0^b-\frac{1}{2}(\bar{L}_{+1}^a\bar{L}_{-1}^b+\bar{L}_{-1}^a\bar{L}_{+1}^b),
\end{equation}
and
\begin{eqnarray}
Z_{abc}&=&L_{0a}\nabla_bL_{0c}-\frac{1}{2}(L_{+1a}\nabla_bL_{-1c}+L_{-1a}\nabla_bL_{+1c}),\nonumber\\
\bar{Z}_{abc}&=&\bar{L}_{0a}\nabla_b\bar{L}_{0c}-\frac{1}{2}(\bar{L}_{+1a}\nabla_b\bar{L}_{-1c}+\bar{L}_{-1a}\nabla_b\bar{L}_{+1c}).
\end{eqnarray}
Apparently $H$ and $\bar{H}$ are symmetric tensor fields and a
straightforward calculation further gives
\begin{equation}
H^{ab}=\bar{H}^{ab}=\frac{1}{4}g^{ab}.
\end{equation}
Concerning $Z$ and $\bar{Z}$ fields, we firstly notice that they are
antisymmetric with respect to the last two indices due to the
Killing equation. On the other hand, it follows from
$\nabla_bH_{ac}=\nabla_b\bar{H}_{ac}=0$ that they are also
antisymmetric with respect to the first and third indices. Therefore
$Z$ and $\bar{Z}$ are actually totally antisymmetric tensor fields
and should be proportional to the three dimensional volume element.
Furthermore, we have
\begin{eqnarray}
\nabla^aZ_{abc}&=&L_{0a}\nabla^a\nabla_bL_{0c}-\frac{1}{2}(L_{+1a}\nabla^a\nabla_bL_{-1c}+L_{-1a}\nabla^a\nabla_bL_{+1c})\nonumber\\
&=&{R_{cb}}^{ad}[L_{0a}L_{0d}-\frac{1}{2}(L_{+1a}L_{-1d}+L_{-1a}L_{+1d})]=0.
\end{eqnarray}
Here we have used the Killing equation in the first step and the
identity
\begin{equation}
\nabla_a\nabla_b\xi_c={R_{cba}}^d\xi_d \label{second}
\end{equation}
for any Killing field $\xi$ in the second step. In addition, in the
last step we have employed the fact that Riemann tensor satisfies
$R_{abcd}=R_{ab[cd]}$. Likewise, we also have
$\nabla^a\bar{Z}_{abc}=0$. Whence the proportional coefficients in
front of the volume element should be constant. In particular, the
explicit calculation yields
\begin{equation}
Z_{abc}=\frac{1}{4}\epsilon_{abc},\bar{Z}_{abc}=-\frac{1}{4}\epsilon_{abc},
\end{equation}
which is consistent with the prevalent claim made in the previous
literature, namely, the two sets of $SL(2,R)$ Lie algebra have the
opposite chirality.

\section{$SL(2,R)$ symmetry for various fields in the BTZ black hole\label{derivation}}
\subsection{Tensor fields}
As a warm-up, let us start with the scalar field $\phi$, whose
equation of motion is given by
\begin{equation}
(\nabla_a\nabla^a-m^2)\phi=0.
\end{equation}
By definition, the Lie derivative acting on the scalar field gives
\begin{equation}
\mathcal{L}_X\mathcal{L}_Y\phi=X^a\nabla_a(Y^b\nabla_b\phi)=(X^a\nabla_aY^b)\nabla_b\phi+X^aY^b\nabla_a\nabla_b\phi.
\end{equation}
Whence it is easy to show
\begin{equation}
\mathcal{L}^2\phi={{Z^a}_a}^b\nabla_b\phi+H^{ab}\nabla_a\nabla_b\phi=\frac{1}{4}g^{ab}\nabla_a\nabla_b\phi=\frac{m^2}{4}\phi.
\end{equation}
Similarly, we have
\begin{equation}
\bar{\mathcal{L}}^2\phi=\frac{m^2}{4}\phi.
\end{equation}
Now let us move onto the massive vector field $A$ with equation of
motion given by
\begin{equation}
{\epsilon_a}^{bc}\nabla_bA_c=-mA_a.
\end{equation}
Whereby we can obtain
\begin{equation}
\nabla_aA^a=-\frac{1}{m}\epsilon^{abc}\nabla_a\nabla_bA_c=-\frac{1}{m}\epsilon^{abc}{R_{abc}}^dA_d=-\frac{1}{m}\epsilon^{abc}{R_{[abc]}}^dA_d=0,
\end{equation}
where the cyclic identity $R_{[abc]d}=0$ has been used in the last
step. On the other hand, we have
\begin{eqnarray}
m^2A^d&=&-m\epsilon^{dea}\nabla_eA_a=\epsilon^{dea}\nabla_e({\epsilon_a}^{bc}\nabla_bA_c)=\epsilon^{ade}{\epsilon_a}^{bc}\nabla_e\nabla_bA_c
\nonumber\\
&=&(g^{dc}g^{eb}-g^{db}g^{ec})\nabla_e\nabla_bA_c=\nabla_a\nabla^aA^d-\nabla_a\nabla^dA^a
\nonumber\\
&=&\nabla_a\nabla^aA^d+\nabla^d\nabla_aA^a-\nabla_a\nabla^dA^a=\nabla_a\nabla^aA^d+R^{dabc}A_cg_{ab}
\nonumber\\
&=&\nabla_a\nabla^aA^d-R^{dc}A_c.
\end{eqnarray}
Acting on this vector field by the Lie derivative, we have
\begin{eqnarray}
\mathcal{L}_X\mathcal{L}_YA_a&=&X^b\nabla_b(\mathcal{L}_YA_a)+\mathcal{L}_YA_b\nabla_aX^b
\nonumber\\
&=&X^b\nabla_b(Y^c\nabla_cA_a+A_c\nabla_aY^c)+(Y^c\nabla_cA_b+A_c\nabla_bY^c)\nabla_aX^b
\nonumber\\
&=&(X^b\nabla_bY^c)\nabla_cA_a+X^bY^c\nabla_b\nabla_cA_a+(X^b\nabla_aY^c)\nabla_bA_c+A_cX^b\nabla_b\nabla_aY^c\nonumber\\
&&+(Y^c\nabla_aX^b)\nabla_cA_b+A_c\nabla_b(Y^c\nabla_aX^b)-A_cY^c\nabla_b\nabla_aX^b.
\end{eqnarray}
Whence we can obtain
\begin{eqnarray}
\mathcal{L}^2A_a&=&{{Z^b}_b}^c\nabla_cA_a+H^{bc}\nabla_b\nabla_cA_a+2{{Z^c}_a}^b\nabla_cA_b+A_c\nabla_b{{Z^c}_a}^b+A^cR_{cabd}H^{bd}-A_cR_{ad}H^{dc}
\nonumber\\
&=&\frac{1}{4}g^{bc}\nabla_b\nabla_cA_a-\frac{1}{2}{\epsilon_a}^{cb}\nabla_cA_b-\frac{1}{4}R_{ac}A^c
\nonumber\\
&=&\frac{1}{4}(m^2+2m)A_a,
\end{eqnarray}
where we have used the identity (\ref{second}) in the first step.
Due to the opposite chirality, we have
\begin{equation}
\bar{\mathcal{L}}^2A_a=\frac{1}{4}(m^2-2m)A_a.
\end{equation}
We conclude this subsection by involving ourselves into the massive
graviton field $h$. The equation of motion is given by
\begin{equation}
{\epsilon_a}^{bc}\nabla_bh_{cd}=-mh_{ad}.
\end{equation}
From this equation, it is easy to find $g^{ab}h_{ab}=0$. In
addition, we can obtain
\begin{eqnarray}
\nabla^ah_{ad}&=&-\frac{1}{m}\epsilon^{abc}\nabla_a\nabla_bh_{cd}=-\frac{1}{m}\epsilon^{abc}({R_{abc}}^eh_{ed}+{R_{abd}}^eh_{ce})
\nonumber\\
 &=&-\frac{1}{m}\epsilon^{abc}(\delta_a^eg_{bd}-g_{ad}\delta_b^e)h_{ce}=-\frac{1}{m}\epsilon^{abc}(g_{bd}h_{ca}-g_{ad}h_{cb})=0.
\end{eqnarray}
Furthermore, we have
\begin{eqnarray}
m^2{h^e}_d&=&-m\epsilon^{efa}\nabla_fh_{ad}=\epsilon^{efa}\nabla_f({\epsilon_a}^{bc}\nabla_bh_{cd})=\epsilon^{aef}{\epsilon_a}^{bc}\nabla_f\nabla_bh_{cd}
\nonumber\\
&=&(g^{ec}g^{fb}-g^{eb}g^{fc})\nabla_f\nabla_bh_{cd}=\nabla_b\nabla^b{h^e}_d-\nabla^c\nabla^eh_{cd}\nonumber\\
&=&\nabla_b\nabla^b{h^e}_d+\nabla^e\nabla^ch_{cd}-\nabla^c\nabla^eh_{cd}\nonumber\\
&=&\nabla_b\nabla^b{h^e}_d-R^{ef}h_{fd}-{R^{eaf}}_dh_{af}.
\end{eqnarray}
Now acting on this massive graviton field, the Lie derivative yields
\begin{eqnarray}
\mathcal{L}_X\mathcal{L}_Yh_{ab}&=&X^c\nabla_c\mathcal{L}_Yh_{ab}+2\mathcal{L}_Yh_{cb}\nabla_aX^c\nonumber\\
&=&X^c\nabla_c(Y^d\nabla_dh_{ab}+2h_{db}\nabla_aY^d)+2(Y^d\nabla_dh_{cb}+h_{db}\nabla_cY^d+h_{cd}\nabla_bY^d)\nabla_aX^c\nonumber\\
&=&(X^c\nabla_cY^d)\nabla_dh_{ab}+X^cY^d\nabla_c\nabla_dh_{ab}+2(X^c\nabla_aY^d)\nabla_ch_{db}+2h_{db}X^c\nabla_c\nabla_aY^d\nonumber\\
&&+2(Y^d\nabla_aX^c)\nabla_dh_{cb}+2h_{db}\nabla_c(Y^d\nabla_aX^c)-2h_{db}Y^d\nabla_c\nabla_aX^c\nonumber\\
&&+2h_{cd}\nabla_b(Y^d\nabla_aX^c)-2h_{cd}Y^d\nabla_b\nabla_aX^c,
\end{eqnarray}
where the symmetrization between the indices $a$ and $b$ is
implicitly assumed for convenience. Whence we can obtain
\begin{eqnarray}
\mathcal{L}^2h_{ab}&=&{{Z^c}_c}^d\nabla_dh_{ab}+H^{cd}\nabla_c\nabla_dh_{ab}+4{{Z^c}_a}^d\nabla_ch_{db}+2h_{db}\nabla_c{{Z^d}_a}^c+2h_{cd}\nabla_b{{Z^d}_a}^c\nonumber\\
&&+2h_{db}{R^d}_{ace}H^{ce}-2h_{db}R_{ae}H^{ed}-2h_{cd}{R^c}_{abe}H^{de}\nonumber\\
&=&\frac{1}{4}g^{cd}\nabla_c\nabla_dh_{ab}-{\epsilon_a}^{cd}\nabla_ch_{db}-\frac{1}{2}{R_a}^eh_{eb}-\frac{1}{2}R_{aceb}h^{ce}\nonumber\\
&=&\frac{1}{4}(m^2h_{ab}+4mh_{ab}-{R_a}^eh_{eb}-R_{aceb}h^{ce})\nonumber\\
&=&\frac{1}{4}[m^2h_{ab}+4mh_{ab}+2\delta_a^eh_{eb}-(g_{ab}g_{ce}-g_{ae}g_{cb})h^{ce}]\nonumber\\
&=&\frac{1}{4}(m^2+4m+3)h_{ab}.
\end{eqnarray}
By the same token, we have
\begin{equation}
\bar{\mathcal{L}}^2h_{ab}=\frac{1}{4}(m^2-4m+3)h_{ab}.
\end{equation}
\subsection{Spinor field}
Let us start with Dirac equation
\begin{equation}
(\gamma^a\nabla_a+m)\psi=0.
\end{equation}
Here $\gamma^a=e_I^a\gamma^I$ and
$\nabla_a=\partial_a+\frac{1}{4}\omega_{IJ a}\gamma^{IJ}$, where
$e_I^a$ form a set of orthogonal normal vector bases, and Gamma
matrices satisfy $\{\gamma^I,\gamma^J\}=2\eta^{IJ}$ with the spin
connection $\omega_{IJa}=e_{Ib}\nabla_ae_J^b$ and
$\gamma^{IJ}=\frac{1}{2}[\gamma^I,\gamma^J]$. Acting on both sides
of Dirac equation with $\gamma^b\nabla_b-m$, we have
\begin{eqnarray}
0&=&(\gamma^b\nabla_b-m)(\gamma^a\nabla_a+m)\psi=(\gamma^a\gamma^b\nabla_a\nabla_b-m^2)\psi\nonumber\\
&=&(g^{ab}\nabla_a\nabla_b+\gamma^{ab}\nabla_a\nabla_b-m^2)\psi=(\nabla_a\nabla^a-m^2)\psi+\gamma^{ab}\nabla_{[a}\nabla_{b]}\psi,
\end{eqnarray}
where $\gamma^{ab}=e_I^ae_J^b\gamma^{IJ}$. To proceed, we notice
\begin{eqnarray}
\nabla_{[a}\nabla_{b]}\psi&=&\partial_{[a}\nabla_{b]}\psi-{\Gamma^c}_{[ab]}\nabla_c\psi+\frac{1}{4}\omega_{IJ[a}\gamma^{IJ}\nabla_{b]}\psi\nonumber\\
&=&\partial_{[a}\partial_{b]}\psi+\frac{1}{4}[\partial_{[a}(\omega_{MNb]}\gamma^{MN}\psi)+\omega_{IJ[a}\gamma^{IJ}\partial_{b]}\psi+\frac{1}{4}\omega_{IJ[a}\gamma^{IJ}\omega_{MNb]}\gamma^{MN}\psi]\nonumber\\
&=&\frac{1}{4}[(\partial_{[a}\omega_{MNb]})\gamma^{MN}\psi+\omega_{IJ[b}\gamma^{IJ}\partial_{a]}\psi+\omega_{IJ[a}\gamma^{IJ}\partial_{b]}\psi+\frac{1}{4}\omega_{IJ[a}\gamma^{IJ}\omega_{MNb]}\gamma^{MN}\psi]\nonumber\\
&=&\frac{1}{4}[(\partial_{[a}\omega_{MNb]})\gamma^{MN}\psi+\frac{1}{4}\omega_{IJ[a}\omega_{MNb]}\gamma^{IJ}\gamma^{MN}\psi]\nonumber\\
&=&\frac{1}{4}\{(\partial_{[a}\omega_{MNb]})\gamma^{MN}\psi+\frac{1}{8}\omega_{IJa}\omega_{MNb}[\gamma^{IJ},\gamma^{MN}]\}\nonumber\\
&=&\frac{1}{4}\{(\partial_{[a}\omega_{MNb]})\gamma^{MN}\psi+\frac{1}{4}\omega_{IJa}\omega_{MNb}(\eta^{JM}\gamma^{IN}-\eta^{JN}\gamma^{IM}-\eta^{IM}\gamma^{JN}+\eta^{IN}\gamma^{JM})\nonumber\\
&=&\frac{1}{4}[(\partial_{[a}\omega_{MNb]})\gamma^{MN}\psi+\omega_{MIa}{\omega^I}_{Nb}\gamma^{MN}\psi]=\frac{1}{4}(\partial_{[a}\omega_{MNb]}+\omega_{MI[a}{\omega^I}_{Nb]})\gamma^{MN}\psi\nonumber\\
&=&\frac{1}{8}R_{abMN}\gamma^{MN}\psi,
\end{eqnarray}
where $\Gamma$ is the Christoffel symbol and the second Cartan
equation has been used in the last step with
$R_{abMN}=R_{abcd}e_M^ce_N^d$. With this observation, we end up with
\begin{equation}
(\nabla_a\nabla^a-m^2+\frac{1}{8}R_{abcd}\gamma^{ab}\gamma^{cd})\psi=0.
\end{equation}
Note that the Lie derivative acting on spinor fields is given by
\begin{equation}
\mathcal{L}_X\psi=X^a\nabla_a\psi-\frac{1}{4}\gamma^{ab}\psi\nabla_bX_a.
\end{equation}
Thus we have
\begin{eqnarray}
\mathcal{L}_X\mathcal{L}_Y\psi&=&X^a\nabla_a\mathcal{L}_Y\psi-\frac{1}{4}\gamma^{ab}\mathcal{L}_Y\psi\nabla_bX_a\nonumber\\
&=&X^a\nabla_a(Y^c\nabla_c\psi-\frac{1}{4}\gamma^{cd}\psi\nabla_dY_c)-\frac{1}{4}\gamma^{ab}(Y^c\nabla_c\psi-\frac{1}{4}\gamma^{cd}\psi\nabla_dY_c)\nabla_bX_a\nonumber\\
&=&(X^a\nabla_aY^c)\nabla_c\psi+X^aY^c\nabla_a\nabla_c\psi-\frac{1}{4}\gamma^{cd}\psi X^a\nabla_a\nabla_dY_c-\frac{1}{4}(X^a\nabla_dY_c)\gamma^{cd}\nabla_a\psi\nonumber\\
&&-\frac{1}{4}(Y^c\nabla_bX_a)\gamma^{ab}\nabla_c\psi+\frac{1}{16}\gamma^{ab}\gamma^{cd}\psi\nabla_d(Y_c\nabla_bX_a)-\frac{1}{16}\gamma^{ab}\gamma^{cd}\psi
Y_c\nabla_d\nabla_bX_a.\nonumber\\
\end{eqnarray}
Whence it is not hard to show
\begin{eqnarray}
\mathcal{L}^2\psi&=&{{Z^a}_a}^c\nabla_c\psi+H^{ac}\nabla_a\nabla_c\psi-\frac{1}{4}\gamma^{cd}\psi
R_{cdae}H^{ae}-\frac{1}{2}{Z^a}_{dc}\gamma^{cd}\nabla_a\psi\nonumber\\
&&+\frac{1}{16}\gamma^{ab}\gamma^{cd}\psi\nabla_dZ_{cba}-\frac{1}{16}\gamma^{ab}\gamma^{cd}\psi R_{abde}{H^e}_c\nonumber\\
&=&\frac{1}{4}(\nabla_a\nabla^a\psi-\frac{1}{2}{\epsilon^a}_{bc}\gamma^{cb}\nabla_a\psi+\frac{1}{16}R_{abcd}\gamma^{ab}\gamma^{cd}\psi)\nonumber\\
&=&\frac{1}{4}[(m^2-\frac{1}{16}R_{abcd}\gamma^{ab}\gamma^{cd})\psi-\frac{1}{2}{\epsilon^a}_{bc}\gamma^{cb}\nabla_a\psi]\nonumber\\
&=&\frac{1}{4}[(m^2+\frac{1}{8}\gamma_{cd}\gamma^{cd})\psi+\frac{1}{2}{\epsilon^a}_{bc}\gamma^{bc}\nabla_a\psi]\nonumber\\
&=&\frac{1}{4}(m^2-\frac{3}{4}+m)\psi,
\end{eqnarray}
where the identity special to three dimension
$\gamma^{ab}=\epsilon^{abc}\gamma_c$ has been used in the last step.
Similarly, we have
\begin{equation}
\bar{\mathcal{L}}^2\psi=\frac{1}{4}(m^2-\frac{3}{4}-m)\psi.
\end{equation}
\section{Quasi-normal modes in the BTZ black hole}
As a recapitulation, we find that the solutions to the equations of
motion for various fields fall into the various representations of
$SL(2,R)$ Lie algebra labeled by the value of the Casimir, i.e.,
\begin{equation}
\mathcal{L}^2\Phi=\lambda_+\Phi,\bar{\mathcal{L}}^2\Phi=\lambda_-\Phi,
\end{equation}
where $\lambda_\pm=\frac{m^2}{4}$ for the scalar field,
$\lambda_\pm=\frac{m^2\pm 2m}{4}$ for the vector field,
$\lambda_\pm=\frac{m^2\pm 4m+3}{4}$ for the tensor field, and
$\lambda_\pm=\frac{m^2\pm m-\frac{3}{4}}{4}$ for our spinor field.
With this observation, the quasi-normal modes can be constructed by
the standard algebraic approach. Speaking specifically, we start
from the highest weight mode which obeys the condition as follows
\begin{equation}
\mathcal{L}^2\Phi_+^{(0)}=\lambda_+\Phi_+^{(0)},
\mathcal{L}_{L_{+1}}\Phi_+^{(0)}=0,
\mathcal{L}_{L_0}\Phi_+^{(0)}=w_+\Phi_+^{(0)},\label{rightweight}
\end{equation}
or
\begin{equation}
\bar{\mathcal{L}}^2\Phi_-^{(0)}=\lambda_-\Phi_-^{(0)},
\mathcal{L}_{\bar{L}_{+1}}\Phi_-^{(0)}=0,
\mathcal{L}_{\bar{L}_0}\Phi_-^{(0)}=w_-\Phi_-^{(0)},\label{leftweight}
\end{equation}
where
\begin{equation}
\Phi_\pm^{(0)}=e^{-i\omega_\pm^0\tau+ip\phi}\Psi_\pm^{(0)}(\rho)\label{expansion}
\end{equation}
with
$\mathcal{L}_{L_0}\Psi_\pm(\rho)=\mathcal{L}_{\bar{L}_0}\Psi_\pm(\rho)=0$\footnote{This
sort of expansion can always be achieved. In particular,
$\Psi_\pm^{(0)}(\rho)$ can be regarded as the coordinate component
of the tensor fields in the coordinate system $\{u,v,\rho\}$. For
our spinor field, it denotes the component associated with the
choice of the orthogonal normal bases as
$e_0^a=\frac{1}{\sinh(\rho)}(\frac{\partial}{\partial\tau})^a$,$e_1^a=\frac{1}{\cosh(\rho)}(\frac{\partial}{\partial\varphi})^a$,
and $e_2^a=(\frac{\partial}{\partial\rho})^a$.}. This implies that
$\omega_\pm^0$ and $p$ correspond to the frequency and angular
momentum of this highest weight mode respectively. By the definition
of quasi-normal mode, we require the imaginary part of
$\omega_\pm^0$ to be negative and $p$ to be real.

Now the resultant quasi-normal modes can be constructed as the
infinite tower of descendant modes, i.e.,
\begin{equation}
(\mathcal{L}_{\bar{L}_{-1}}\mathcal{L}_{L_{-1}})^n\Phi_\pm^{(0)}
\end{equation}
with $n=0,1,2,\cdot\cdot\cdot$. Employing the commutation relation
(\ref{commutation}), one can show that the conformal weight is given
by
\begin{equation}
w_+=\frac{1\pm\sqrt{1+4\lambda_+}}{2},w_-=\frac{1\pm\sqrt{1+4\lambda_-}}{2},
\end{equation}
and the corresponding quasi-normal frequencies can be worked out as
\begin{equation}
\omega_\pm^n=\pm p-2i(w_\pm+n).
\end{equation}
As expected, the result is in good agreement with the previous
calculation\cite{BSS,SS,CL}.
\section{Conclusion}
Instead of solving the equations of motion analytically, we have
constructed the quasi-normal modes for various fields in the BTZ
black hole and determined its frequencies in a uniform way by
invoking the algebraic approach. The result is in good agreement
with the previous calculation as it should be. To achieve this, the
primary task is to show the solutions to equation of motion fall
into the representation of the $SL(2,R)$ Lie algebra, which is
fulfilled by the explicit tensor and spinor analysis without
resorting to any specific coordinate system. To make such an
analysis as simple as possible, we have introduced two tensor fields
intrinsic to the $SL(2,R)$ Casimir of Killing fields and unveiled
their intriguing features by relating them to the metric and volume
element respectively. As shown, these two tensor fields conspire to
play an important role in organizing and simplifying the relevant
tensor and spinor analysis.

We conclude with some generalizations of our work in various
directions.  Firstly, although the analysis for more general fields
are expected to go straightforward, it is interesting to investigate
how the quasinormal frequencies are quantitatively related to the
mass parameter appearing in the equation of motion. On the other
hand, besides the BTZ black hole considered here, there are other
somewhat complicated three dimensional black holes such as warped
black holes and self-dual warped black holes\cite{ALPSS,CMN}. It is
intriguing to show how these cases can be fitted into our framework
such that the relevant spacetime symmetry can be made transparent
and the whole calculation can be simplified similarly\cite{Zhang1}.
In addition, associated with the near horizon geometry, the
$SL(2,R)$ symmetry plays an important role in the context of
Kerr/CFT correspondence\cite{GHSS}. Thus it is rewarding to see
whether our strategy is also applicable to this higher dimensional
spacetime by doing something like $3+1$ decomposition\cite{Zhang2}.
Finally, it is definitely worthwhile to explore whether our strategy
can be extended to the more challenging case of hidden conformal
symmetry\cite{CMS}.
\section*{Acknowledgements}
The author is grateful to Bin Chen for his talk at KITPC, which
sparks his dive into this project. In addition, he would like to
take this opportunity to thank Ronggen Cai and Elias Kiritsis for
their very help to make possible his attending the long term AdS/CFT
programme at KITPC. This research was supported in part by the PKIP
of Chinese Academy of Sciences with Grant No. KJCX2.YW.W10. It was
also supported by a European Union grant
FP7-REGPOT-2008-1-CreteHEPCosmo-228644.
\section*{Appendices}
\subsection*{A Calculation of $Z$ and $\bar{Z}$ fields}
By the fact that $\nabla_a\xi_b=\frac{1}{2}(d\xi)_{ab}$ for any
Killing field $\xi$, we have
\begin{eqnarray}
Z_{abc}=Z_{[abc]}&=&\frac{1}{2}[L_{0[a}dL_{0bc]}-\frac{1}{2}(L_{+1[a}dL_{-1[bc]}+L_{-1[a}dL_{+1bc]})]\nonumber\\
&=&\frac{1}{6}[L_{0}\wedge dL_{0abc}-\frac{1}{2}(L_{+1}\wedge
dL_{-1abc}+L_{-1}\wedge dL_{+1abc})],
\end{eqnarray}
where
\begin{eqnarray}
L_{0a}&=&g_{ab}L_{0}^b=-\frac{1}{4}(du)_a+\frac{\cosh(2\rho)}{4}(dv)_a,\nonumber\\
L_{-1a}&=&g_{ab}L_{-1}^b=e^{-u}[\frac{\sinh(2\rho)}{4}(dv)_a-\frac{1}{2}(d\rho)_a],\nonumber\\
L_{+1a}&=&g_{ab}L_{+1}^b=e^{u}[\frac{\sinh(2\rho)}{4}(dv)_a+\frac{1}{2}(d\rho)_a],
\end{eqnarray}
and
\begin{eqnarray}
(dL_0)_{ab}&=&\frac{\sinh(2\rho)}{2}(d\rho)_a\wedge(dv)_b, \nonumber\\
(dL_{-1})_{ab}&=&e^{-u}[-\frac{\sinh(2\rho)}{4}(du)_a\wedge(dv)_b+\frac{\cosh(2\rho)}{2}(d\rho)_a\wedge(dv)_b+\frac{1}{2}(du)_a\wedge(d\rho)_b], \nonumber\\
(dL_{+1})_{ab}&=&e^u[\frac{\sinh(2\rho)}{4}(du)_a\wedge(dv)_b+\frac{\cosh(2\rho)}{2}(d\rho)_a\wedge(dv)_b+\frac{1}{2}(du)_a\wedge(d\rho)_b].
\end{eqnarray}
With this, we can finally obtain
\begin{equation}
Z=\frac{\sinh(2\rho)}{16}du\wedge dv\wedge d\rho.
\end{equation}
By the symmetry between $u$ and $v$, we have
\begin{equation}
\bar{Z}=\frac{\sinh(2\rho)}{16}dv\wedge du\wedge
d\rho=-\frac{\sinh(2\rho)}{16}du\wedge dv\wedge d\rho.
\end{equation}
\subsection*{B A little bit of Clifford algebra}
Firstly by the identity
\begin{equation}
[A,BC]=\{A,B\}C-B\{A,C\},
\end{equation}
we have
\begin{equation}
[\gamma^I,\gamma^M\gamma^N]=2(\eta^{IM}\gamma^N-\eta^{IN}\gamma^M),
\end{equation}
which further gives
\begin{equation}
[\gamma^I,\gamma^{MN}]=2(\eta^{IM}\gamma^N-\eta^{IN}\gamma^M).
\end{equation}
Next by the Jacobi identity, we have
\begin{eqnarray}
[\gamma^{IJ},\gamma^{MN}]&=&\frac{1}{2}[[\gamma^I,\gamma^J],\gamma^{MN}]=\frac{1}{2}([\gamma^I,[\gamma^J,\gamma^{MN}]]-[\gamma^J,[\gamma^I,\gamma^{MN}])\nonumber\\
&=&2(\eta^{JM}\gamma^{IN}-\eta^{JN}\gamma^{IM}-\eta^{IM}\gamma^{JN}+\eta^{IN}\gamma^{JM}).
\end{eqnarray}
\subsection*{C A little bit of spinor analysis}
Associated with the choice of the orthogonal normal bases as
$e_0^a=\frac{1}{\sinh(\rho)}(\frac{\partial}{\partial\tau})^a$,$e_1^a=\frac{1}{\cosh(\rho)}(\frac{\partial}{\partial\varphi})^a$,
and $e_2^a=(\frac{\partial}{\partial\rho})^a$, the non-vanishing
spin connection is given by
\begin{equation}
\omega_{02a}=-\omega_{20a}=-\frac{\cosh(\rho)}{2}[(du)_a+(dv)_a],
\omega_{12a}=-\omega_{21a}=\frac{\sinh(\rho)}{2}[(du)_a-(dv)_a].
\end{equation}
Thus we have
\begin{eqnarray}
\mathcal{L}_{L_0}\Psi(\rho)&=&-(\frac{\partial}{\partial
u})^a[\partial_a+\frac{1}{2}(\omega_{02a}\gamma^{02}+\omega_{12a}\gamma^{12})]\Psi(\rho)+\frac{\sinh(2\rho)}{8}\gamma^{ab}\Psi(\rho)(d\rho)_a(dv)_b\nonumber\\
&=&\frac{1}{4}[\cosh(\rho)\gamma^{02}-\sinh(\rho)\gamma^{12}]\Psi(\rho)+\frac{\sinh(2\rho)}{8}\gamma^{IJ}\Psi(\rho)e_I^ae_J^b(d\rho)_a(dv)_b\nonumber\\
&=&\frac{1}{4}[\cosh(\rho)\gamma^{02}-\sinh(\rho)\gamma^{12}]\Psi(\rho)+\frac{\sinh(2\rho)}{8}[\frac{1}{\sinh(\rho)}\gamma^{20}-\frac{1}{\cosh(\rho)}\gamma^{21}]\Psi(\rho)\nonumber\\
&=&0.
\end{eqnarray}
Similarly, we can obtain $\mathcal{L}_{\bar{L}_0}\Psi(\rho)=0$.

\end{document}